# High upper critical field (120 T) with small anisotropy of highly hydrogen-substituted SmFeAsO epitaxial film

Kota Hanzawa,[1] Jumpei Matsumoto,[1] Soshi Iimura,[2,3,4] Yoshimitsu Kohama,[5] Hidenori Hiramatsu,[1,4] and Hideo Hosono[2,4]

(1) Laboratory for Materials and Structures, Institute of Innovative Research, Tokyo Institute of Technology, Yokohama, Kanagawa 226-8503, Japan.

(2) National Institute for Materials Science (NIMS), Tsukuba, Ibaraki 305-004, Japan.

(3) PRESTO, Japan Science and Technology Agency, Kawaguchi, Saitama 332-0012, Japan.

(4) Materials Research Center for Element Strategy, Tokyo Institute of Technology, Yokohama, Kanagawa 226-8503, Japan.

(5) Institute for Solid State Physics, The University of Tokyo, Kashiwa, Chiba 277-8581, Japan.

**Abstract**

The electronic transport properties of a highly hydrogen-substituted 1111-type SmFeAsO epitaxial film with high critical-temperature ($T_c$ = 45 K) were investigated under high magnetic fields. By using a single-turn magnet generating up to 130 T, we clarified that the upper critical field ($\mu_0 H_{c2}$) of SmFeAsO$_{0.65}$H$_{0.35}$ is 120 T at the nearly low-temperature limit of 2.2 K for $\mu_0 H \parallel ab$. The angular dependence of $\mu_0 H_{c2}$ revealed that the anisotropic parameter ($\gamma$) around $T_c$ is ~2, which is comparable with that of a practical candidate 122-type BaFe$_2$As$_2$ with lower $T_c$ and much smaller than that of F-substituted SmFeAsO. The small $\gamma$ mainly originates from the high hydrogen incorporation. The extremely high $\mu_0 H_{c2}$ and small $\gamma$, together with the high $T_c$ and high critical current density, suggest that SmFeAsO$_{1-x}$H$_x$ has high potential for the superconducting electromagnets and cables.

**Introduction**

The upper critical field ($\mu_0H_{c2}$) is one of the most important properties for evaluation of superconductors and their applications. The analysis of $\mu_0H_{c2}$ provides an understanding of the basic characteristics for superconducting applications, such as the anisotropic parameter ($\gamma \equiv H_{c2\|}/H_{c2\perp}$, where the directions of the applied external magnetic fields are parallel ($\mu_0H \|$) and perpendicular ($\mu_0H \perp$) to the sample surface), and clarifies the mechanism of Cooper-pair breaking. Therefore, $\mu_0H_{c2}$ has been extensively investigated for various superconductors, including high critical-temperature ($T_c$) superconductors. For next-generation superconductor applications, such as electromagnets and cables, not only high $T_c$ and high critical current density ($J_c$), but also high $\mu_0H_{c2}$ and small $\gamma$ are required. However, simultaneous realization of these properties is challenging. For instance, MgB$_2$ with a high $T_c$ of 40 K possesses a small $\gamma$ of ~2 [1], while $\mu_0H_{c2\|}$ at 0 K ($\mu_0H_{c2\|}(0)$) is as low as 18 T [2]. In the case of cuprates, such as YBa$_2$Cu$_3$O$_{7-\delta}$ and Bi$_2$Sr$_2$Ca$_2$Cu$_3$O$_x$ that have high $T_c$ values of 40 and >90 K, the large $\gamma$ of above, for example, 50 for Bi$_2$Sr$_2$Ca$_2$Cu$_3$O$_x$ [3], is undesirable for applications, although $\mu_0H_{c2\|}(0)$ is very high (>200 T) [4].

Iron-based superconductors may be suitable for electromagnet and cable applications [5] because of their high $T_c$ [6] and three-dimensional electronic structures [7] despite their layered crystal structures. 122-Type K-substituted BaFe$_2$As$_2$ exhibits a high $T_c$ of 38 K [8] and nearly isotropic $\mu_0H_{c2}$ [9], which has been inferred to reach 70 T. Co-substituted BaFe$_2$As$_2$ also induces superconductivity with $T_c$ = 22 K, and $\mu_0H_{c2}$ is known to be isotropic around 0 K [10, 11], where it is presumed that the Pauli paramagnetic limit causes the pair breaking and contributes to the isotropic $\mu_0H_{c2}$. Because of these outstanding superconducting properties, including high $J_c$, 122-type

BaFe$_2$As$_2$ is regarded as the most promising candidate among iron-based superconductors for practical applications. This has led to research on its potential for tape [12] and wire materials [13].

Among iron-based bulk superconductors, the highest $T_c$ of 55 K has been recorded for the 1111-type superconductor, whose crystal structure consists of alternatively stacked FeAs and $RE$O ($RE$ = rare earth metal) layers (see **Fig. 1(a)**). The superconductivity emerges by electron doping through partial substitution of F$^-$ in the O$^{2-}$ sites of SmFeAsO [14]. The high $T_c$ of 55 K and possible high $\mu_0H_{c2}(0)$ of >200 T [15, 16] are favorable for superconducting applications. However, research on 1111-type superconductors has progressed less than that of other iron-based superconductors, such as 122-type superconductors. One reason is the lack of a complete electronic phase diagram for 1111-type superconductors. The low solubility of F$^-$ in the O$^{2-}$ site (0.2e$^-$ per Fe) has restricted investigation of the superconducting properties in the overdoped region. Another reason is the difficulty in preparing $RE$FeAsO. Only a few research groups have achieved single-crystal growth [17] or epitaxial thin-film fabrication for F substitution [18], even though the 122- and 11-systems are easily obtained.

Our research group has proposed the hydride ion (H$^-$) as an alternative to F$^-$ to generate electrons through O$^{2-}$-site substitution [19–21]. By the H substitution, the electron concentration increased by up to 0.8e$^-$ per Fe. This increase led to discovery of a top-hat-shaped $T_c$ dome between 5% and 50% H substitution and another antiferromagnetic phase in the overdoped region [22, 23]. In addition, we recently developed a H-substitution procedure for SmFeAsO epitaxial thin films using topochemical reaction with binary dihydrides [24, 25]. The relationship between the crystallographic orientation and sample configuration is shown in **Fig. 1(b)**, which indicates that the out-of-plane and in-plane crystallographic orientations were [001] SmFeAsO ∥ [001] MgO and [100] SmFeAsO ∥ [100] MgO, respectively. This method enables substitution of H$^-$ for the O$^{2-}$ site, creating carrier electrons in the 1111-type epitaxial film and realizing high-$T_c$ superconductivity with a high H concentration reaching ~35% of the O sites, which exceeds the solubility limit of F doping (<20%). The effectiveness of this procedure has been validated for another $RE$-based 1111-type NdFeAsO epitaxial film [26–28].

Importantly, H-substituted SmFeAsO films exhibit not only high $T_c$, but also high $J_c$ of 1 MA/cm$^2$ in the self-field and ~0.5 MA/cm$^2$ under $\mu_0H$ = 9 T at 2 K [25]. This $J_c$ value is comparable with those of 122-type high-$J_c$ films [12]. Furthermore, the unique electronic structure of H-substituted SmFeAsO may be advantageous for superconducting applications. In the F-substituted 1111-type superconductor, the insulator $RE$O layer suppresses interaction between the conducting FeAs layers [29], which contributes to increase in the anisotropy of $\mu_0H_{c2}$ ($\gamma$ > 4). In contrast, in the H-substituted 1111-type superconductor, the As 4p and H 1s orbitals in the interlayer are covalently bonded, and they should make the Fermi surface partially three-dimensional [30]. Such an interaction does not appear in the F-substituted system. Thus, reduction of the anisotropy of $\mu_0H_{c2}$ can be expected for the H-substituted sample.

In this study, we investigated the pair-breaking mechanism and anisotropy of a highly H-substituted SmFeAsO epitaxial film under high magnetic fields. Owing to the high $\mu_0H_{c2}(0)$, a single-turn magnet was used for experimental investigation in high magnetic fields of up to 130 T. We found that H-substituted SmFeAsO exhibits high $\mu_0H_{c2}(0)$ (reaching 120 T) when the external magnetic field is applied along the $ab$ plane. Additionally, we clarified the small $\gamma$ value of 2.1, which is comparable with that of 122-type BaFe$_2$As$_2$ ($\gamma \leq 2$ [9, 11]) and almost half of that of F-substituted SmFeAsO ($\gamma > 4$ [16]). The small $\gamma$ value of H-substituted SmFeAsO mainly originates from realization of three-dimensional superconductivity owing to high H incorporation.

**Experimental**

**Topochemical H substitution for SmFeAsO film**

The highly H-substituted SmFeAsO epitaxial thin film with a thickness of 80 nm was prepared by topotactic chemical reaction between a pristine undoped SmFeAsO epitaxial film on a MgO (001) single-crystal substrate (10 mm × 10 mm × 0.5 mm (thickness)) and CaH$_2$ powder. The detailed procedure for sample preparation can be found in Ref. [24]. The H concentration in the film was ~35% of the O sites [24]. Results of structure analysis of the SmFeAsO$_{0.65}$H$_{0.35}$ film by X-ray diffraction is summarized in **Fig. S1** [31].

**Characterization of electronic transport properties using high-field magnets**

The temperature and field dependences of the electronic transport properties were characterized by using three different magnets. The experiments in the steady fields were carried out in a Physical Property Measurement System (PPMS, Quantum Design Inc., USA). The experiments in the pulsed fields were performed in the non-destructive and single-turn pulsed magnets at the Institute for Solid State Physics, The University of Tokyo. The maximum field strength generated by the PPMS, non-destructive, and single-turn magnets were 9, 60, and 130 T, respectively. **Figure 1(c)** and **(d)** show the results obtained by using the 9 T PPMS, which were used to estimate the normal resistance of the sample. For the experiment with the 60 T pulsed magnet, the SmFeAsO$_{0.65}$H$_{0.35}$ film was mounted on a rotator probe, where the angle between the field direction and the *c* axis of the epitaxial film was defined as $\theta$. In this measurement, the magnetoresistance of the sample was measured in a four-wire configuration at 50 kHz with a digital lock-in procedure. In the impedance measurement with the single-turn magnet, the field dependence of the impedance was measured at 125 MHz [32]. Note that in the impedance measurement, because we employed a two-wire configuration, the impedance of wires is non-negligible and a constant background appears.

**Results and Discussion**

First, we measured the electronic transport properties of the SmFeAsO$_{0.65}$H$_{0.35}$ epitaxial film with a non-destructive pulsed magnet generating up to 60 T. The main carrier polarity was determined to be electrons by Hall-effect measurements conducted at 50 and 80 K (**Figure S2** [31]), and the carrier density ($n = n_e - n_h$, where $n$, $n_e$, and $n_h$ are whole carrier, electron, and hole densities, respectively) was calculated to be ~5 × 10$^{21}$ cm$^{-3}$ by using a multiband model at high-field limit ($\mu_0 H \to \infty$); i.e., Hall resistance $R_{xy}/\mu_0 H = -1/e(n_e - n_h)t$, where $e$ and $t$ are the elementary charge and film thickness. This carrier density is much higher than that of F-substituted NdFeAsO and comparable with that of H-substituted NdFeAsO [26]. **Figure 1(c)** and **1(d)** are temperature ($T$) dependence of resistance ($R$) of the SmFeAsO$_{0.65}$H$_{0.35}$ epitaxial film as a function of $\mu_0 H$ applied along the $c$ axis and $ab$ plane up to 9 T, where the $T_c$ was 45 K. **Figure 1(e)** and **1(f)** show the field dependence of the $R$ in the H-substituted SmFeAsO epitaxial film for the $\mu_0 H \perp$ and $\mu_0 H \parallel ab$ configurations at $T$ from 2.2 to 50 K, respectively. The magnetoresistance ($MR$) for $\mu_0 H \perp ab$ showed the tendency of saturation above 30 K, while that for $\mu_0 H \parallel ab$ was hardly saturated even around $T_c$. For estimation of $\mu_0 H_{c2}$ at each $T$ from the $R$–$\mu_0 H$ curves, we defined $\mu_0 H^{80}$, where the normal state $R$ ($R_{\text{normal}}$) reaches 80% with correction based on the normal-state MR, that is, $R(\mu_0H^{80})$ = $0.8R_{normal}/(1 + MR)$. Here, the $R_{normal}$ value at each measurement temperature was estimated from linear extrapolation of the zero-field R–T data (black fitting lines in **Fig. 1(c)** and **(d)**). $MR$ (= $(R(\mu_0H) − R(0))/R(0)$, **Fig. S3** [31]) was calculated from the $R$–$\mu_0H$ curves at 50 K in **Fig. 1(e)** and **(f)**. The measurable maximum $\mu_0H_{c2}$ in **Fig. 1(e)** and **(f)** was 50 T for $\mu_0H \perp ab$ at 25 K and 59 T for $\mu_0H \parallel ab$ at 35 K, clearly suggesting that $\mu_0H_{c2}(0)$ was much higher than 60 T, especially for $\mu_0H \parallel ab$.

To experimentally determine $\mu_0H_{c2}(0)$ for $\mu_0H \parallel ab$, we performed radio-frequency (rf) impedance measurement in two-wire configuration with a single-turn magnet up to 130 T. In our impedance analysis, the sample resistance was roughly proportional to the inverse ratio of the amplitude of the reflection rf signal to the incident rf excitation with a field-independent background ($\Gamma^{-1}$). $\Gamma^{-1}$ as a function of $\mu_0H$ at 2.2 K is shown in **Fig. 2(a)**, where the black dashed line indicates the constant background, and the red and blue curves were obtained during the up- and down-sweep processes of the pulsed fields, respectively. Even though $\Gamma^{-1}$ remained constant between 0 and ~100 T due to the wire resistance, $\Gamma^{-1}$ steeply increased at higher fields with small hysteresis between the up (red arrow) and down sweeps (blue arrow). An enlarged view of the high-field region in **Fig. 2(a)** is shown in **Fig. 2(b)**. From the least-squares fits for the up- and down-sweep data (black dashed and solid lines, respectively), clear saturation of $\Gamma^{-1}$ was observed above ~125 T. By taking $\mu_0H$ where $\Gamma^{-1}$ became 80% of the saturated $\Gamma^{-1}$ as $\mu_0H_{c2}$, $\mu_0H_{c2}$ of SmFeAsO$_{0.65}$H$_{0.35}$ at 2.2 K was estimated to be 120 T for $\mu_0H \parallel ab$.

The temperature dependence of $\mu_0H_{c2}$ is shown in **Fig. 3(a)** and **(b)**. **Figure 3(a)** depicts fitting curves using the single-band model based on Werthamer–Helfand–Hohenberg (WHH) theory [33]. If the spin–orbital scattering effect is considered to be negligible, the WHH equation can be simplified to

$$\ln\frac{T_c}{T} = \frac{1}{2}\psi\left[\frac{1}{2}+\frac{2(1+\alpha)T_cH_{c2}}{\pi^2(-dH_{c2}/dT)T}\right] + \frac{1}{2}\psi\left[\frac{1}{2}+\frac{2(1-\alpha)T_cH_{c2}}{\pi^2(-dH_{c2}/dT)T}\right] - \psi\left(\frac{1}{2}\right), \quad (1)$$

where $\psi$, $\alpha$, and $-dH_{c2}/dT$ are the digamma function, Maki parameter, and slope of superconducting to the normal-state phase boundary near $T_c$, respectively. The fitting results in **Fig. 3(a)** correspond to the cases of $\alpha = 0$ (green curve), 0.5 (purple curve), and 1.2 (orange curve). For $\mu_0H \perp ab$, the best fit to the WHH model was obtained with $\alpha = 0.5$, where $\mu_0H_{c2}(0)_\perp$ was estimated to be ~80 T. In contrast, for $\mu_0H \parallel ab$, the theoretical curve reproduced the experimental results with $\alpha = 1.2$. Because $\alpha$ is defined as $\alpha = \sqrt{2}\mu_0H^{orb}(0)/\mu_0H^P(0)$, where $\mu_0H^{orb}(0)$ is the orbital-limiting field, the large $\alpha$ suggests that the Pauli-limiting field ($\mu_0H^P(0)$) is small. This would indicate that the Cooper pair-breaking mechanism is attributed to the Zeeman effect. Moreover, the superconductor–metal transition in **Fig. 2** exhibits first-order phase transition like behavior, that is, hysteresis appears before and after the phase transition. Even though the present system includes a large amount of dopant, the hysteresis may indicate emergence of the Fulde–Ferrell–Larkin–Ovchinnikov state for the $\mu_0H \parallel ab$ configuration [34], which has also been reported for the other iron-based superconductors FeSe [35] and LiFeAs [36], and organic superconductors, such as $\kappa$-(BEDT-TTF)$_2$Cu(NCS)$_2$ [37].

Next, to take into account the multiband electronic structure of SmFeAsO, we fitted the data to the dirty-limit two-band model [38]:

$$\ln\frac{T}{T_c} = -\left\{Re\psi\left[\frac{1}{2}+h(i+p)\right] + Re\psi\left[\frac{1}{2}+h(i+p\eta)\right] - 2\psi\left(\frac{1}{2}\right) + \frac{\lambda_0}{w}\right\}/2$$

$$+\left\{\left\{Re\psi\left[\frac{1}{2}+h(i+p)\right] - Re\psi\left[\frac{1}{2}+h(i+p\eta)\right] - \frac{\lambda_-}{w}\right\}^2/4 + \frac{\lambda_{12}\lambda_{21}}{w^2}\right\}^{1/2}, \quad (2)$$

where $i$ and $\lambda_{mn}$ are the imaginary unit and the superconducting coupling constant between the $m$th and $n$th bands, respectively. $\lambda_-$, $\lambda_0$, and $w$ are $\lambda_{11} - \lambda_{22}$, $(\lambda_-  + 4\lambda_{12}\lambda_{21})^{1/2}$, and $\lambda_{11}\lambda_{22} - \lambda_{12}\lambda_{21}$, respectively. $p$ and $\eta$ are the diffusivity ratios: $p = D_1/D_0$ and $\eta = D_2/D_1$, where $D_0$ is the quantum diffusivity and $D_m$ is the electron diffusivity in the $m$th band due to nonmagnetic impurity scattering. $h$ is defined as $D_1\mu_0H_{c2}/2\Phi_0T$, where $\Phi_0$ is the flux quantum. The fitted results are shown in **Fig. 3(b)**. A previous study of SmFeAsO suggested that intraband coupling is dominant rather than interband coupling, and the fit is insensitive to $w$ compared with $D_m$ [16], which is confirmed also in our fittings using different sets of coupling constants. Thus, to simplify the fitting process, we fixed the interband coupling constants to $\lambda_{11} = \lambda_{22} = 0.5$ and the intraband coupling constants to $\lambda_{12} = \lambda_{21} = 0.25$. Consequently, we found that the results can be reproduced with $p_\perp = 2.0$ and $\eta_\perp = 0.53$ for the $\mu_0H \perp ab$ case and with $p_\parallel = 0.75$ and $\eta_\parallel = 0.6$ for the $\mu_0H \parallel ab$ case in the whole $T$ region. Extrapolation of the fitting curve to zero temperature gave $\mu_0H_{c2}(0) \approx 80$ T for $\mu_0H \perp ab$. Here, because the electron motions along the $ab$ plane and $c$ axis are affected by the Lorentz force for the $\mu_0H \perp$ and $\mu_0H \parallel ab$ configurations, $p_\perp$ and $p_\parallel$ are expressed as $p_\perp = D_{1\parallel}/D_0$ and $p_\parallel = (D_{1\parallel}D_{1\perp})^{1/2}/D_0 = (p_\perp D_{1\perp}/D_0)^{1/2}$, respectively, where $D_{1\parallel}$ and $D_{1\perp}$ are the electron diffusivities in one of the bands along the $ab$ plane and $c$ axis, respectively. The $D_{1\parallel}$ and $D_{1\perp}$ values estimated by the above equations for SmFeAsO$_{0.65}$H$_{0.35}$ are 1.2 and 0.2 cm$^2$/s, respectively. These parameters are more isotropic than those in F-substituted NdFeAsO ($D_{1\parallel}$ =5.8 cm$^2$/s and $D_{1\perp}$ =0.02 cm$^2$/s) [39], indicating that the electron diffusion along the $c$ axis in H-substituted SmFeAsO is strongly enhanced compared with that in the F-substituted sample, although there should be an influence of the rare earth metal. This indicates that H incorporation increases the interlayer interaction in SmFeAsO and likely contributes to the reduction of the anisotropy.

**Figure 3(c)** shows the $R$–$\mu_0H$ curves at 37.5 K with different field angles $\theta$. With increasing $\theta$, the jump in the $R$–$\mu_0H$ curve gradually shifted to the higher field side. The $\theta$ dependence of $\mu_0H_{c2}$ was estimated from the angle-dependent magnetoresistance and fitted to the two-band theoretical model using Eq. (2), where $D_m$ was replaced by the angle-dependent electron diffusivity: $D_m(\theta) = [D_{m\parallel}^2\cos^2\theta + D_{m\parallel}D_{m\perp}\sin^2\theta]^{1/2}$. The fitting curve is shown in **Fig. 3(d)**, and it well reproduces the observed angle dependence, supporting the reliability of the parameters obtained by the two-band model. This experiment revealed that although there is a non-negligible angle dependence in $\mu_0H_{c2}$, the anisotropy of the critical field ($\gamma = H_{c2\parallel}/H_{c2\perp}$) seems to be very small despite the layered structure of H-substituted SmFeAsO, which consists of individual SmO insulating and FeAs conductive layers.

We also estimated the irreversible fields ($\mu_0H_{irr}$) by taking the field where the resistance was 10% of the normal state (**Figure S4** [31]). $\mu_0H_{irr}$ was fitted by a simple power law: $\mu_0H_{irr} = \mu_0H_{irr}(0)(1 - T/T_c)^z$. The fit suggests that $\mu_0H_{irr}(0)$ was extrapolated to 90 and 105 T under $\mu_0H \perp$ and $\mu_0H \parallel ab$, respectively. The anisotropic parameter $\gamma_{irr}$ was also small ($\gamma_{irr} \approx 2$). Furthermore, in a previous study, we found that the SmFeAsO$_{0.65}$H$_{0.35}$ film exhibits high critical current density $J_c$ exceeding 1 MA/cm$^2$ in a self-field and ~0.5 MA/cm$^2$ even with a high field of 9 T at 2 K [25], which is comparable with that of the 122-type iron-based superconductor [12]. These remarkable and unique characteristics under a magnetic field, such as the high $\mu_0H_{c2}(0)$ of 120 T, relatively isotropic $\gamma$ and $\gamma_{irr}$, and potentially high $J_c$, are because of the specific features of H, and they are advantageous for use of SmFeAsO$_{0.65}$H$_{0.35}$ as a material for superconducting applications.

The $\gamma$ values of SmFeAsO$_{0.65}$H$_{0.35}$ obtained in the non-destructive pulsed magnet (red solid circles) are shown in **Fig. 4** along with those of representative iron-based superconductors and other high-$T_c$ superconductors. The experimental $\gamma$ values of SmFeAsO$_{0.65}$H$_{0.35}$ were 2.1 at 35 K, and 2.3 even at 40 K, which is just below the $T_c$ value of 45 K. The $\gamma(T/T_c)$ of SmFeAsO$_{0.65}$H$_{0.35}$ was clearly smaller than those of high-$T_c$ curates [40] and other 1111-type iron-based superconductors [16, 39]. For example, F-substituted SmFeAsO, which has the same parent phase as H-substituted SmFeAsO, exhibits a large $\gamma$ of more than 4 at approximately $T_c$ [16]. Moreover, the $\gamma(T/T_c)$ value is comparable with those of MgB$_2$ [1], and 11- [41, 42], 111- [43], and 122-type iron-based superconductors [9, 10, 44]. Here, it should be noted that the $\gamma$ values of the other iron-based superconductors tend to decrease with decreasing $T$. This indicates that $\gamma$ of H-substituted SmFeAsO would further decrease at lower $T$.

For further comparison, some superconducting properties of representative iron-based superconductors are summarized in **Table 1**, where the orbital-limiting field $\mu_0H_{c2}^{orb}(0)$ was estimated by $\mu_0H_{c2}^{orb}(0) = -0.69(dH_{c2}/dT)T_c$ and the coherence length $\xi(0)$ at 0 K was calculated by $\xi(0)_\parallel = (\Phi_0/2\pi\mu_0H_{c2}^{orb}(0)_\perp)^{1/2}$ and $\xi(0)_\perp = \Phi_0/2\pi\mu_0H_{c2}^{orb}(0)_\parallel\xi(0)_\parallel$. The $\xi(0)_\parallel$ values of SmFeAsO$_{0.8}$F$_{0.2}$ and SmFeAsO$_{0.65}$H$_{0.35}$ are almost the same (~26 Å), whereas $\xi(0)_\perp$ of SmFeAsO$_{0.65}$H$_{0.35}$ (13 Å) is three-times longer than that of SmFeAsO$_{0.8}$F$_{0.2}$ (4.4 Å). The $\xi(0)_\perp$ value of SmFeAsO$_{0.65}$H$_{0.35}$ is much longer than the FeAs layer separation (~8.5 Å), suggesting that the interlayer coupling of the FeAs layers is enhanced and the superconductivity should be three dimensional.

Finally, we discuss the origin of the small $\gamma$ in the SmFeAsO$_{0.65}$H$_{0.35}$ epitaxial film. First, the SmFeAsO$_{0.65}$H$_{0.35}$ epitaxial film has a large amount of incorporated H [24]. The high electron density induced by the high H substitution would lead to delocalization of the charge density within the SmO layer, which would not occur in the F-substituted system owing to the lower electron density. Furthermore, the high impurity content introduces disorder in the system. This disorder enlarges the orbital-limiting field [45, 46], and consequently the isotropic Pauli limiting field would dominate pair breaking. Second, SmFeAsO$_{0.65}$H$_{0.35}$ exhibits large lattice contraction along the $c$ axis. The substitution-induced contraction in the $c$-axis lattice parameter, $\Delta c = c_x - c_0$, where $c_x$ and $c_0$ are the lengths of the $c$ axis in polycrystalline SmFeAsO$_{1-x}$(H/F)$_x$ and undoped SmFeAsO, respectively, is larger than that in the $a$-axis lattice parameter, $\Delta a = a_x - a_0$, in SmFeAsO$_{0.87}$H$_{0.13}$ ($\Delta c/\Delta a = 2.34$) [19]. In contrast, $\Delta c/\Delta a$ of SmFeAsO$_{0.85}$F$_{0.15}$ is 0.93 [47]. This indicates that the lattice of H-substituted SmFeAsO preferentially shrinks along the $c$-axis compared with that of the F-substituted sample, leading to enhancement of the coupling between the FeAs and SmO layers, and likely contributing to covalent bonding between the As 4s and H 1s orbitals, as in CaFeAsH [19, 30]. Third, the single-crystal substrate of the film sample enhances lattice contraction in the H-substituted SmFeAsO film. It has been reported that $\Delta c/\Delta a$ of the SmFeAsO$_{0.65}$H$_{0.35}$ epitaxial film is 3.48 [24], suggesting that epitaxial strain further stimulates lattice shrinkage along the $c$ axis. Hence, we speculate that these factors originating from H substitution and epitaxial growth result in the small $\gamma$ value of H-substituted SmFeAsO.

**Conclusions**

We have investigated the upper critical field ($\mu_0 H_{c2}$) of the highly H-substituted SmFeAsO$_{0.65}$H$_{0.35}$ epitaxial thin film with $T_c = 45$ K under high magnetic fields of up to 130 T generated by three different magnets (superconducting, non-destructive, and single-turn magnets). The electronic transport measurements using the non-destructive magnet with the maximum magnetic field of 60 T suggest that $\mu_0 H_{c2}$ with external fields along the $c$ axis can be extrapolated to 80 T at 0 K based on both the WHH and two-band models.

Through measurement of $\mu_0H_{c2\parallel}$ using strong magnetic fields of up to 130 T generated by the single-turn magnet within the *ab* plane, we clarified that $\mu_0H_{c2\parallel}$ of SmFeAsO$_{0.65}$H$_{0.35}$ reaches 120 T at 2.2 K. The experimentally confirmed critical field should be almost the same as its low-temperature limit $\mu_0H_{c2}(0)$, and it is the highest value among those of other iron-based superconductors. Furthermore, compared with the anisotropic parameters of $\mu_0H_{c2}$ of 1111-type iron-based superconductors, including F-substituted SmFeAsO ($\gamma > 4$), the anisotropic parameter of SmFeAsO$_{0.65}$H$_{0.35}$ is remarkably small ($\gamma \approx 2.1$). We infer that the small $\gamma$ mainly originates from the three-dimensionality enhanced by high-density H incorporation and its correlated epitaxial strain, which would lead to strong interaction between the FeAs and SmO layers and enhancement of the interlayer coupling of the FeAs layers. Owing to the high and nearly isotropic $\mu_0H_{c2}$, as well as the high $T_c$ and high $J_c$, H-substituted SmFeAsO is a promising candidate for superconducting applications, such as the wires and tapes for cables and high-field electromagnets, even though improvement of the H-substitution process or development of novel H-substitution methods should be necessary for fabrication of long-length technological conductors.

**Acknowledgments**

This work was supported by the Ministry of Education, Culture, Sports, Science, and Technology (MEXT) through the Element Strategy Initiative to Form Core Research Center (Grant Number JPMXP0112101001). The experiments in the pulsed fields were carried out under the Visiting Researcher's Program of the Institute for Solid State Physics, The University of Tokyo (approval No. B254). K. H. was supported by JSPS KAKENHI Grant Numbers JP20K15170 and JP 22K14601.

**Figures and Table**

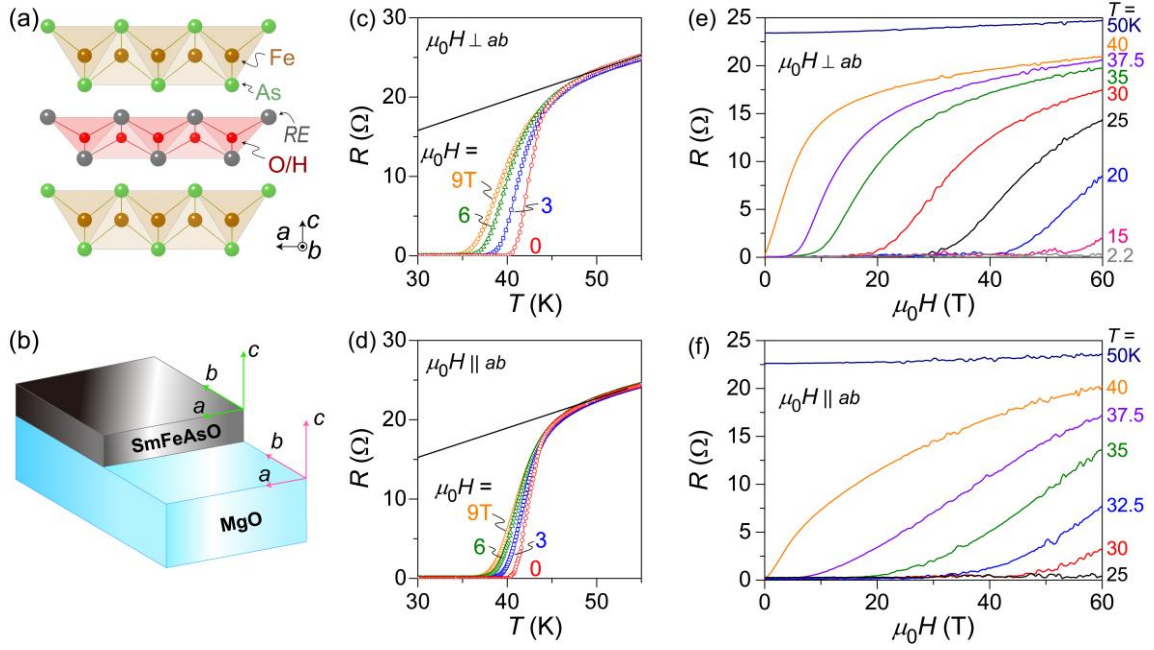

**Figure 1.** Sample configuration and electronic transport properties of the H-substituted SmFeAsO epitaxial film. (**a**) Crystal structure of 1111-type $RE$FeAsO with space group $P4/nmm$, where $RE$ is a rare earth metal, such as Sm. $H^-$ partially occupies the $O^{2-}$ sites in the $RE$O layer. (**b**) Relationship between the crystallographic orientation and the sample configuration of the H-substituted SmFeAsO epitaxial film on a MgO single-crystal substrate. (**c**, **d**) $T$ dependence of $R$ of the SmFeAsO$_{0.65}$H$_{0.35}$ epitaxial film as a function of $\mu_0 H$ applied (**c**) perpendicular ($\perp$) and (d) parallel ($\parallel$) to the $ab$ plane of the sample with a superconducting magnet of up to 9 T. The red, blue, green, and orange symbols correspond to $\mu_0 H = 0$, 3, 6, and 9 T, respectively. The black lines are the results of least-squares fitting for the normal conducting states. $\mu_0 H_{c2}$ was estimated using the definition of $T_c^{80}$, at which $R$ decreases to 80% of that in the normal conducting state. (**e**, **f**) Electronic transport properties under 60 T pulsed magnetic fields at $T = 2.2$–50 K. The field was applied (**e**) $\perp$ and (**f**) $\parallel$ $ab$ plane.

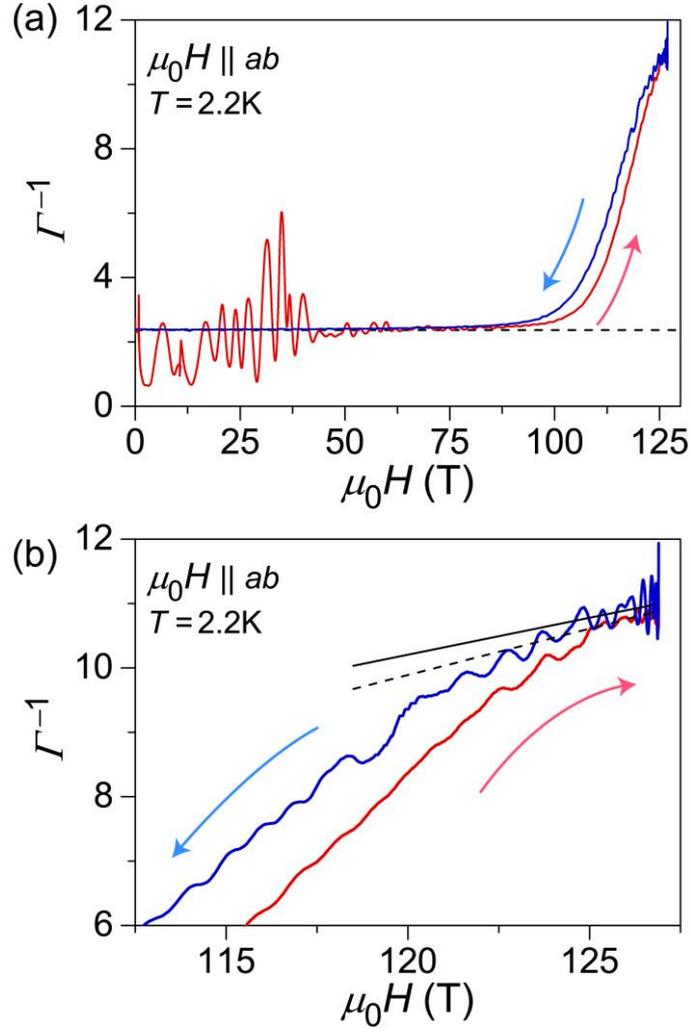

**Figure 2.** Impedance measurement under high magnetic fields of up to 130 T at 2.2 K. The magnetic field was generated by a single-turn magnet and applied within the *ab* plane. The vertical axis is the inverse ratio of the amplitude of the reflection rf signal to the incident rf excitation with constant background ($\Gamma^{-1}$). (**a**) $\Gamma^{-1}$ in the field region from 0 to 130 T. (**b**) Enlarged image of (**a**) in the high-field region. The constant background is indicated by a black dashed line in (**a**). The red and blue arrows denote the directions of the field sweep. The black dashed and solid lines in (**b**) are the least-squares fits of the normal-state resistance in the up and down sweeps.

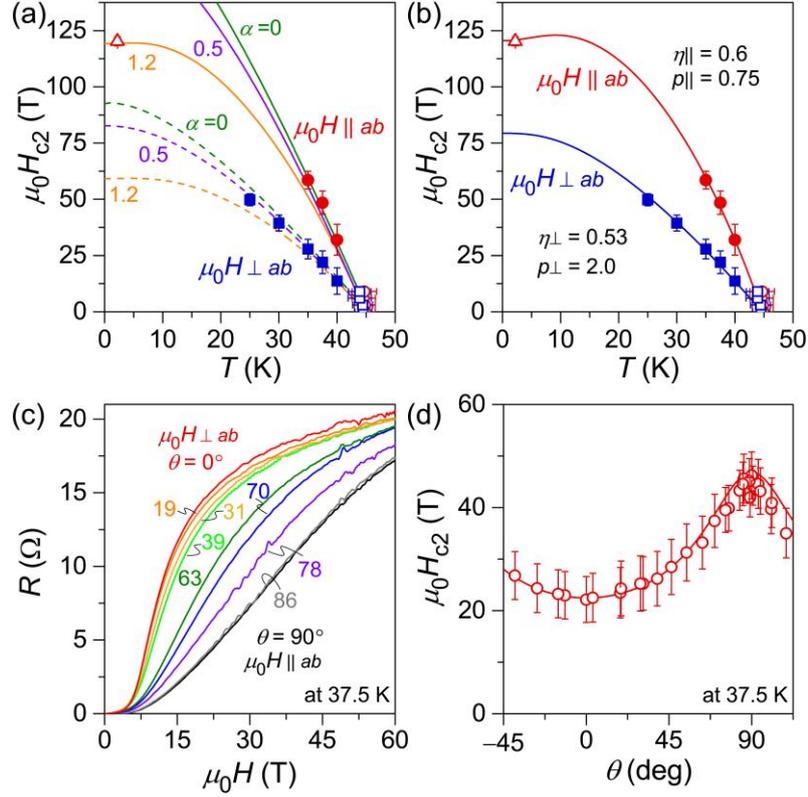

**Figure 3.** The experimentally observed $\mu_0 H_{c2}$ in the H-substituted SmFeAsO epitaxial film and its angular dependence. (**a**, **b**) $\mu_0 H_{c2}$ along with the theoretical fits calculated by the (**a**) WHH and (**b**) two-band models. The red and blue symbols are the experimental data under $\mu_0 H \parallel$ and $\perp ab$. The $\mu_0 H_{c2}$ data were obtained by using superconducting (open circles and open squares), non-destructive pulse (solid circles and solid squares), and single-turn magnets (open triangles). The errors of observed $\mu_0 H_{c2}$ were estimated from $\mu_0 H^{70}$ and $\mu_0 H^{90}$, and indicated by error bars. This error should sufficiently include influences of the frequency in the impedance measurement and difference between impedance and resistance measurements. $\alpha$ in (**a**) represents the Maki parameter in the WHH model. $p$ and $\eta$ in (**b**) are the ratios of the electronic diffusivities in the two-band model. (**c**) $R$–$\mu_0 H$ curves under $\mu_0 H$ of up to 60 T at 37.5 K as a function of the angle ($\theta$). (**d**) Angular dependence of $\mu_0 H_{c2}$ at 37.5 K. The red curve is the fit obtained by two-band analysis.

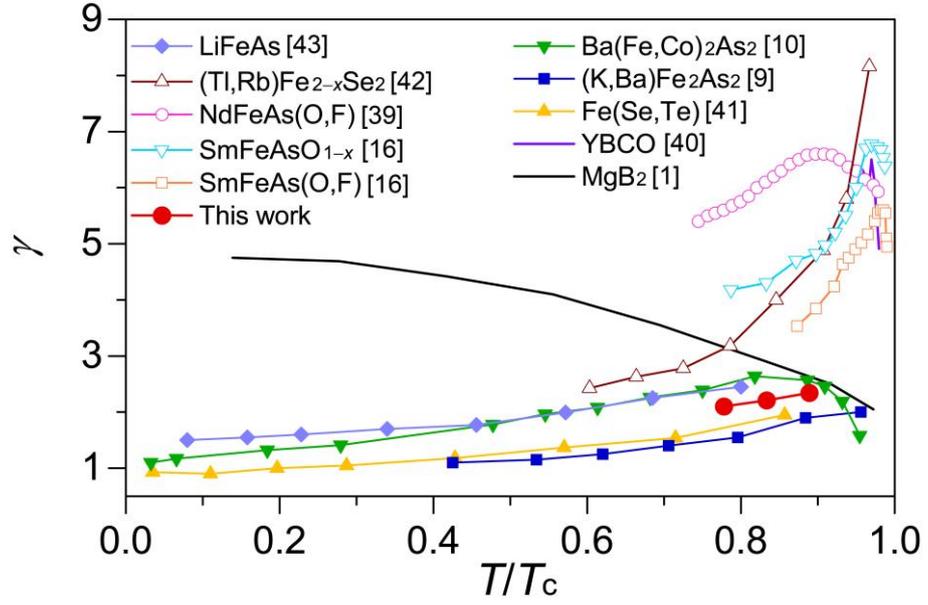

**Figure 4.** Temperature dependences of anisotropic parameter ($\gamma$) for SmFeAsO$_{0.65}$H$_{0.35}$ and other iron-based [9, 10, 16, 39, 41–43] and high-$T_c$ superconductors [1, 40].

**Table 1.** Superconducting parameters of representative iron-based superconductors. $\mu_0H_{c2}$ is the experimentally observed $\mu_0H_{c2}$, where $\mu_0H_{c2}$ without an observation temperature corresponds to the low-temperature limit $\mu_0H_{c2}(0)$. $\gamma(0)$ and $\gamma(0.9)$ are the anisotropic parameters ($\gamma(T/T_c)$) at $T/T_c = 0$ and 0.9, respectively.

| Compounds | $T_c$ (K) | $\mu_0H_{c2}^{\|ab}$ (T) | $\mu_0H_{c2}^{\|c}$ (T) | $\mu_0H_{c2}^{orb}(0)^{\|ab}$ (T) | $\mu_0H_{c2}^{orb}(0)^{\|c}$ (T) | $\gamma(0)$ | $\gamma(0.9)$ | $\xi(0)^{\|ab}$ (Å) | $\xi(0)^{\|c}$ (Å) | Refs. |
|---|---|---|---|---|---|---|---|---|---|---|
| Fe(Se,Te) | 14 | 42 | 43 | 86 | 37 | 0.95 | 2 | 30 | 13 | 41 |
| (Rb,Tl)Fe$_{2-y}$Se$_2$ | 33 | 55 (18 K) | 60 | 273 | 45 | | 4.9 | 27 | 4.5 | 42 |
| LiFeAs | 18 | 24.2 | 15 | 40 | 14.5 | 1.5 | 2.5 | 48 | 17 | 43 |
| Ba(Fe,Co)$_2$As$_2$ | 22 | 55 | 50 | 58 | 35 | 1.1 | 2.5 | 26 | 25 | 10, 44 |
| (K,Ba)Fe$_2$As$_2$ | 38 | 57 (10 K) | 55 (9 K) | 104 | 56 | | 1.9 | 24 | 13 | 9 |
| NdFeAs(O,F) | 55 | 53 (35 K) | 45 (16 K) | 250-300 | 38-150 | | 6.6 | 23 | 2.6 | 39 |
| SmFeAsO$_{1-x}$ | 55 | 52 (44 K) | 57 (28 K) | 380 | 84 | | 4.9 | 20 | 4.4 | 16 |
| SmFeAs(O,F) | 55 | 58 (32 K) | 46 (8 K) | 280 | 47 | | 4 | 26 | 4.4 | 16 |
| SmFeAs(O,H) | 55 | 120 | 50 (25 K) | 186 | 93 | | 2.3 | 26.5 | 13 | This work |

**Supplemental Material for "High upper critical field (120 T) with small anisotropy of highly hydrogen-substituted SmFeAsO epitaxial film"**

Kota Hanzawa,[1] Jumpei Matsumoto,[1] Soshi Iimura,[2,3,4] Yoshimitsu Kohama,[5] Hidenori Hiramatsu,[1,4] and Hideo Hosono[2,4]

(1) Laboratory for Materials and Structures, Institute of Innovative Research, Tokyo Institute of Technology, Yokohama, Kanagawa 226-8503, Japan.

(2) National Institute for Materials Science (NIMS), Tsukuba, Ibaraki 305-004, Japan.

(3) PRESTO, Japan Science and Technology Agency, Kawaguchi, Saitama 332-0012, Japan.

(4) Materials Research Center for Element Strategy, Tokyo Institute of Technology, Yokohama, Kanagawa 226-8503, Japan.

(5) Institute for Solid State Physics, The University of Tokyo, Kashiwa, Chiba 277-8581, Japan.

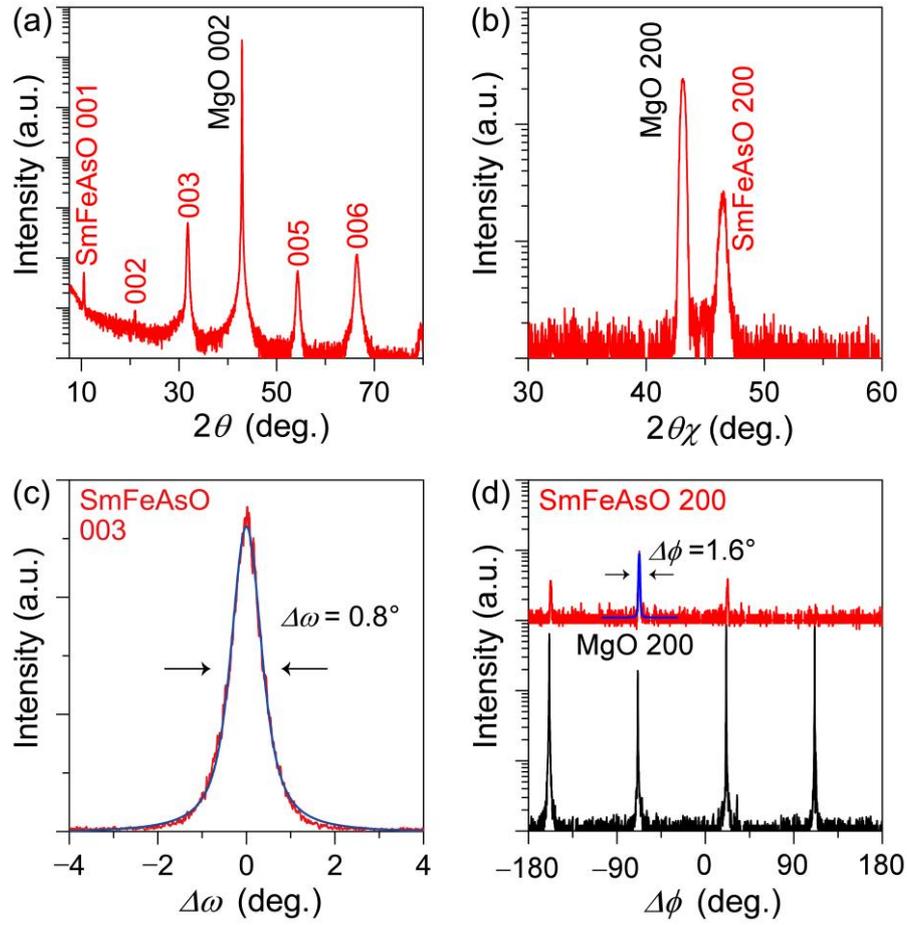

**Figure S1.** Structure analysis of the 80-nm-thick SmFeAsO$_{0.65}$H$_{0.35}$ epitaxial thin film grown on a MgO (001) single-crystal substrate. (**a**) Out-of-plane and (**b**) in-plane XRD patterns. X-ray rocking curves at the SmFeAsO$_{0.65}$H$_{0.35}$ (**c**) 003 and (**d**) 200 diffractions. The blue curves in (**c**) and (**d**) are the results of fitting.

**Text S1.**

The crystal structure of the H-substituted SmFeAsO film was analyzed by X-ray diffraction (XRD). The out-of-plane (**Figure S1(a)**) and in-plane (**Figures S1(b)** and **(d)**) XRD patterns guarantee preservation of the epitaxial relationship between the SmFeAsO film and a MgO single-crystal (10 mm × 10 mm × 0.5 mm (thickness)) without segregation of any impurity phases via the topotactic reaction. From the out-of-plane and in-plane diffractions, the lattice parameters for the *a* and *c* axes were estimated to be 3.903 and 8.438 Å, respectively. The full width at half maxima of the out-of-plane ($\Delta\omega$, **Fig. S1(c)**) and in-plane ($\Delta\phi$, **Fig. S1(d)**) X-ray rocking curve measurements were 0.8° and 1.6°. These results ensure the consistency of the prepared H-substituted SmFeAsO film with a ~35% H-substituted SmFeAsO epitaxial film in a previous study, where *a* = 3.903 Å, *c* = 8.488 Å, $\Delta\omega$ = 0.6°, and $\Delta\phi$ = 1.2° [24].

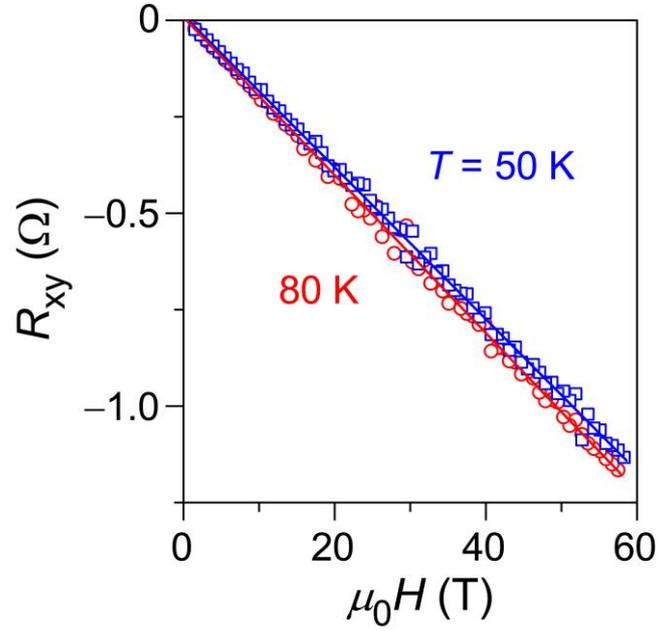

**Figure S2.** Hall-effect measurements of the SmFeAsO$_{0.65}$H$_{0.35}$ epitaxial film using a pulsed-field magnet at 50 (blue squares) and 80 K (red circles). The Hall voltage was measured along the in-plane (*ab* plane) direction. $\mu_0H$ was applied along the out-of-plane direction, that is, the *c* axis. The lines are the results of least-squares fitting.

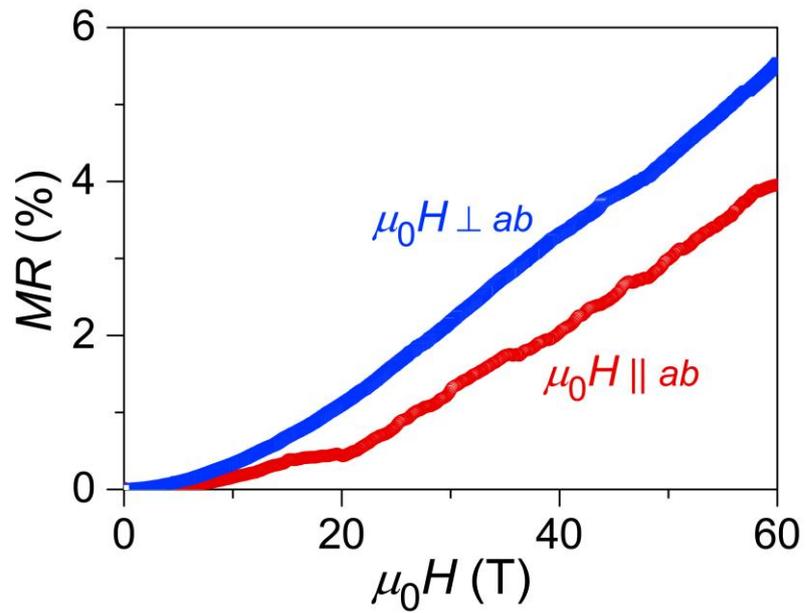

**Figure S3.** Magnetoresistance ($MR = (R(\mu_0H) - R(0))/R(0)$) calculated from the $R$–$\mu_0H$ curves at 50 K in **Figs. 1(e)** and **1(f)** for $\mu_0H \perp$ (blue) and $\mu_0H \parallel ab$ (red).

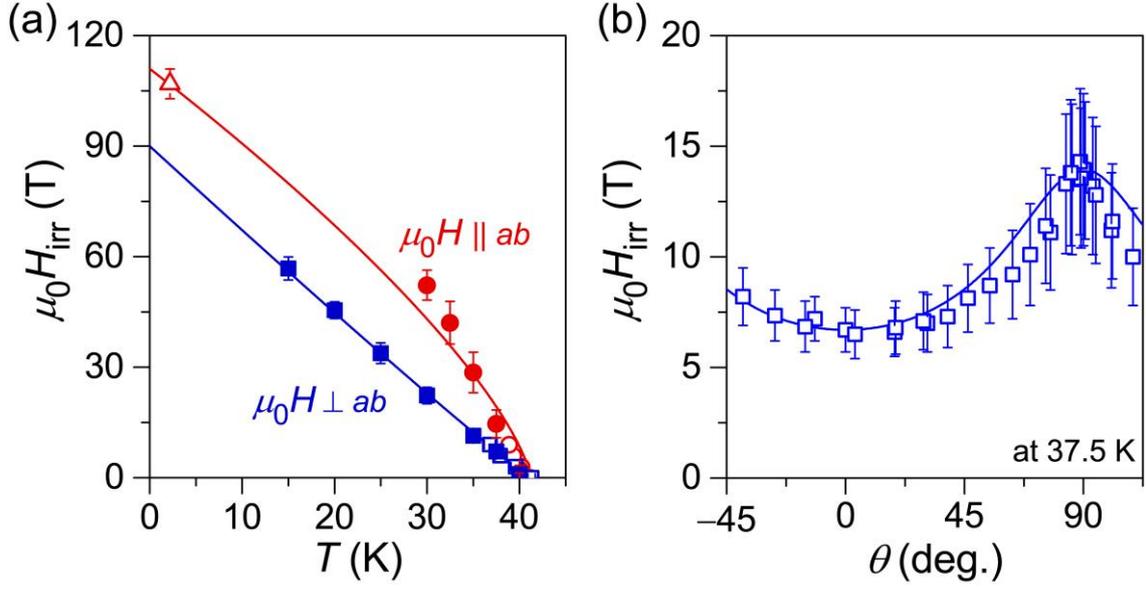

**Figure S4.** Irreversible field ($\mu_0 H_{irr}$) of the SmFeAsO$_{0.65}$H$_{0.35}$ epitaxial film. (**a**) $T$ dependence of $\mu_0 H_{irr}$. $\mu_0 H_{irr}$ was roughly estimated from $\mu_0 H$, where $R$ becomes 10% of that of the normal state. The open circles and open squares, solid circles and solid squares, and open triangle correspond to the experimental data in the superconducting, non-destructive, and single-turn pulse magnets, respectively. The red and blue curves are the fitting curves based on a simple power law: $\mu_0 H_{irr} = \mu_0 H_{irr}(0)(1 - T/T_c)^z$, where $z = 0.73$ and 1.05, respectively. (**b**) Angular ($\theta$) dependence of $\mu_0 H_{irr}$ at 37.5 K, where the blue curve is from an equation expressing the anisotropy of $\mu_0 H_{irr}$: $\mu_0 H_{irr} = \mu_0 H_{irr}(0)/(\cos^2\theta + \gamma\sin^2\theta)^{1/2}$.